# Development of a novel MPGD-based drift chamber for the NSCL/FRIB S800 spectrometer

**M. Cortesi,**[a,*] **J. Pereira,**[a] **D. Bazin,**[a,b] **Y. Ayyad,**[a] **G. Cerizza,**[a] **R. Fox,**[a] **and R.G.T. Zegers**[a,b,c]

[a] *National Superconducting Cyclotron Laboratory (NSCL),*
*Michigan State University (MSU),*
*East Lansing, Michigan 48824, USA*

[b] *Department of Physics and Astrophysics,*
*Michigan State University (MSU)*
*East Lansing, Michigan 48824, USA*

[c] *Joint Institute for nuclear Astrophysics – Center for the Evolution of the Elements,*
*Michigan State University (MSU)*
*East Lansing, Michigan 48824, USA*
*E-mail*: cortesi@nscl.msu.edu

ABSTRACT: The performance of a novel tracking detector developed for the focal plane of the NSCL/FRIB S800 magnetic spectrometer is presented. The detector comprises a large-area drift chamber equipped with a hybrid Micro-Pattern Gaseous Detector (MPGD)-based readout. The latter consists of a position-sensitive Micromegas detector preceded by a two-layer M-THGEM multiplier as a pre-amplification stage. The signals from the Micromegas readout are processed by a data acquisition system based on the General Electronics for TPC (GET). The drift chamber has an effective area of around 60x30 cm$^2$, which matches to the very large acceptance of the S800 spectrometer. This work discusses in detail the results of performance evaluation tests carried out with a low-energy alpha-particles source and with high-energy heavy-ion beams with the detector installed at the S800 focal plane. In this latter case, the detector was irradiated with a 150 MeV/u $^{78}Kr^{36+}$ beam as well as a heavy-ion fragmentation cocktail beam produced by the $^{78}Kr^{36+}$ beam impinging on a thin beryllium target. Sub-millimeter position resolution is obtained in both dispersive and non-dispersive directions.

KEYWORDS: Heavy-ion detector; Particle tracking detectors (Gaseous detectors); Spectrometers.

[*] Corresponding author.

## Contents

## 1. Introduction

The S800 superconducting spectrometer [1] is used for studying nuclear reactions induced by radioactive beams with energy between 10 and 100 MeV/u. It is in operation at National Superconducting Cyclotron Laboratory (NSCL) since the end of the ‘90s. Radioactive beams are produced at the NSCL Coupled-Cyclotron Facility (CCF) [1], [2] and delivered to the S800 experimental area through the A1900 fragment separator [3], which filters the beam to select one of a few specific isotopes of interest. The spectrometer was designed for high-precision measurements of scattering angles within ±2 msr and momentum $p/\Delta p=2\times10^4$, combined to a large solid angle (20 msr) and a large momentum acceptance (6%) [1]. The large solid angle and the high-resolution (1/10,000) were optimized for magnetic rigidities of up to 4 Tm. The S800 has been an indispensable apparatus for the broad physics program of the NSCL with fast rare isotope beams, being the most heavily-used experimental device at NSCL. The S800 spectrograph will continue to serve the nuclear physics/astrophysics community for experiments with rare isotope beams also during FRIB operation.

A crucial component for the performance of the S800 spectrometer is the focal plane detector system, which consists of an array of various detector technologies for trajectory reconstruction as well as particle identification (PID). The basic configuration of the focal plane includes two x/y drift chambers for tracking, an ionization chamber for atomic number identification by energy loss measurement, and a plastic scintillator for timing (as well as energy loss). Downstream of the plastic scintillator, a CsI(Na) hodoscope is deployed to identify atomic charge states of the implanted nuclei via total kinetic energy (TKE) measurement. Additionally, the hodoscope is also used to detect gamma-rays from isomeric states.

The tracking system provides a measurement of the transverse positions and angles of charged particles on an event-by-event basis, with 100% particle transmission. It consists of two large-area Drift Chambers (DCs) placed 1 meter apart, which provide a position resolution of 0.5 mm (σ). Under these conditions, the angle of the particle trajectory can be determined with a resolution of better than 2 mrad, even when multiple scattering effects in the tracking detector are considered.

Presently the S800 tracking system is based on the Cathode Readout Drift Counters (CRDCs) technology [2]. The operational principle of the CRDC is similar to the one of a single-wire drift detector [4], except that the position along the wire is derived by reading out the induced charge on a segmented cathode.

In this work the operational mechanism and performance of a novel tracking detector planned for the upgrade of the S800 focal plane is described for the first time. The design of the new drift chamber (DC) is based on the old CRDC concept, with the exception of the anode readout which is based on a hybrid Micro-Pattern Gaseous Detector readout. Performance evaluations under irradiation with small lab source (a few MeV alpha–particle emitted by a 228-Th source) as well as with heavy-ion beams will be presented and discussed in detail. In the latter case the detector was irradiated by a $^{78}Kr^{36+}$ beam of around 150 MeV/u, as well as by a cocktail of heavy-ions beam produced by the $^{78}Kr$ nuclei impinging on a Be target.

## 2. The MPGD-based drift chamber

A schematic view of the drift chamber and details of the gas avalanche readout assembly can be seen in figure 1. Most of the detector components, including windows frames and field cage for the drift region, are based on the CRDC original design and are explicitly described in ref. [2], with the exception of the readout scheme that replaces the single-wire structure. The new readout comprises a position-sensitive Micromegas [5] board preceded by a two-layer M-THGEM [6] as a pre-amplification stage (figure 2).

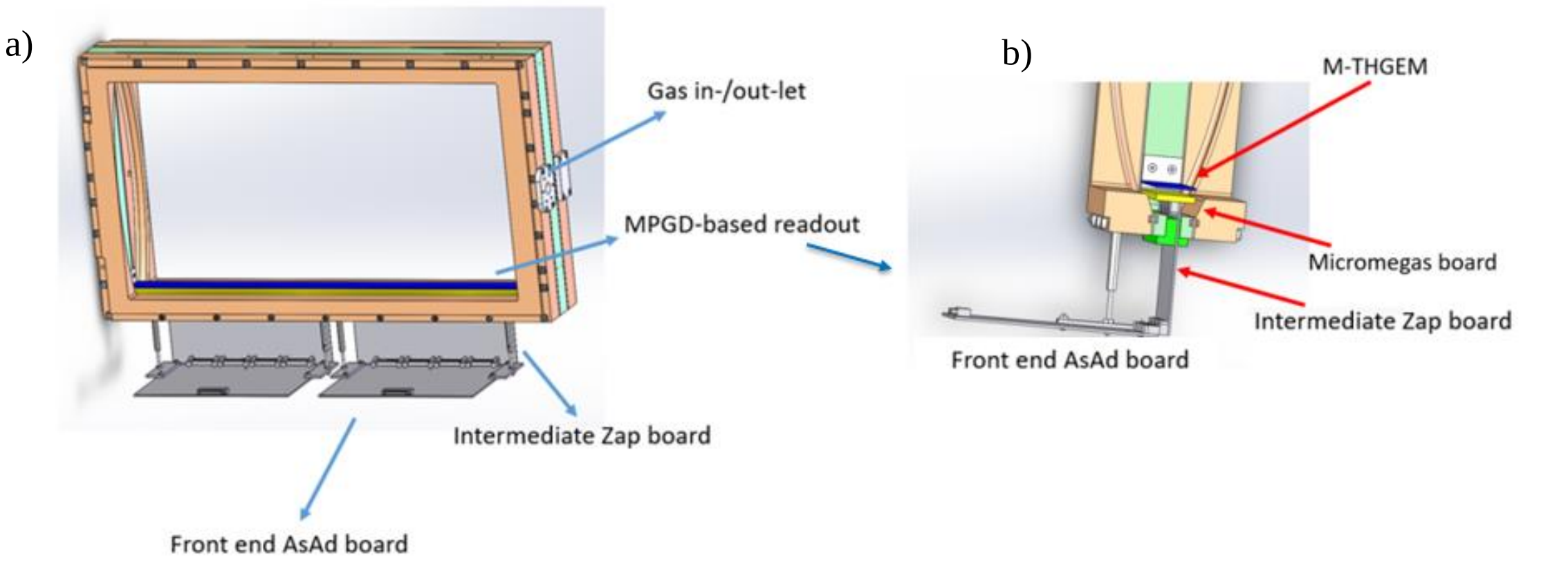

**Figure 1.** Technical drawing of the detector assembling (front view in part a), and detailed of the position-sensitive readout installed in the inner detector frame and associated electronics (side view cut in part b).

The Micromegas readout consists of a micro-mesh (Bopp 45/18) stretched over an anode board segmented into 480 rectangular pads. Each pad has an effective area of 28 mm × 1.125 mm (figure 3), with a distance between neighboring pads of 100 μm. Small insulating pillars support the micromesh on top of the board and define a 128 μm thick amplification gap. The Micromegas board has a rectangular shape, with dimensions of 3 cm and 62 cm along the beam direction and the dispersive coordinate, respectively. Signals are induced on the pads by the fast collection of electrons created during the avalanche process in the gas gap, and by the retrograde motion of avalanche ions. The signals are routed out through intermediate PCB (ZAP) boards that host passive elements for discharge protection. They are further processed by the AsAD (ASIC Support & Analog-Digital conversion) front-end electronics boards, as shown in figure 1 (see Section 2.1 for more details), directly connected to the ZAP board.

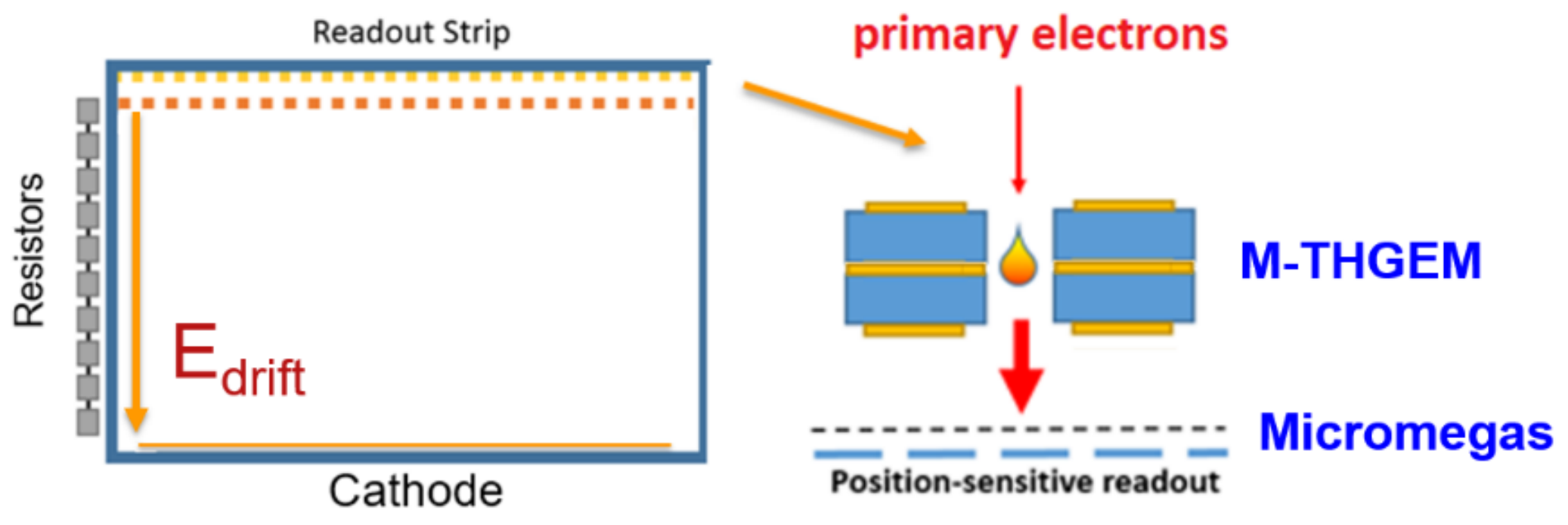

**Figure 2.** Schematic drawing of the novel drift chamber design and of the hybrid Micromegas/M-THGEM avalanche readout.

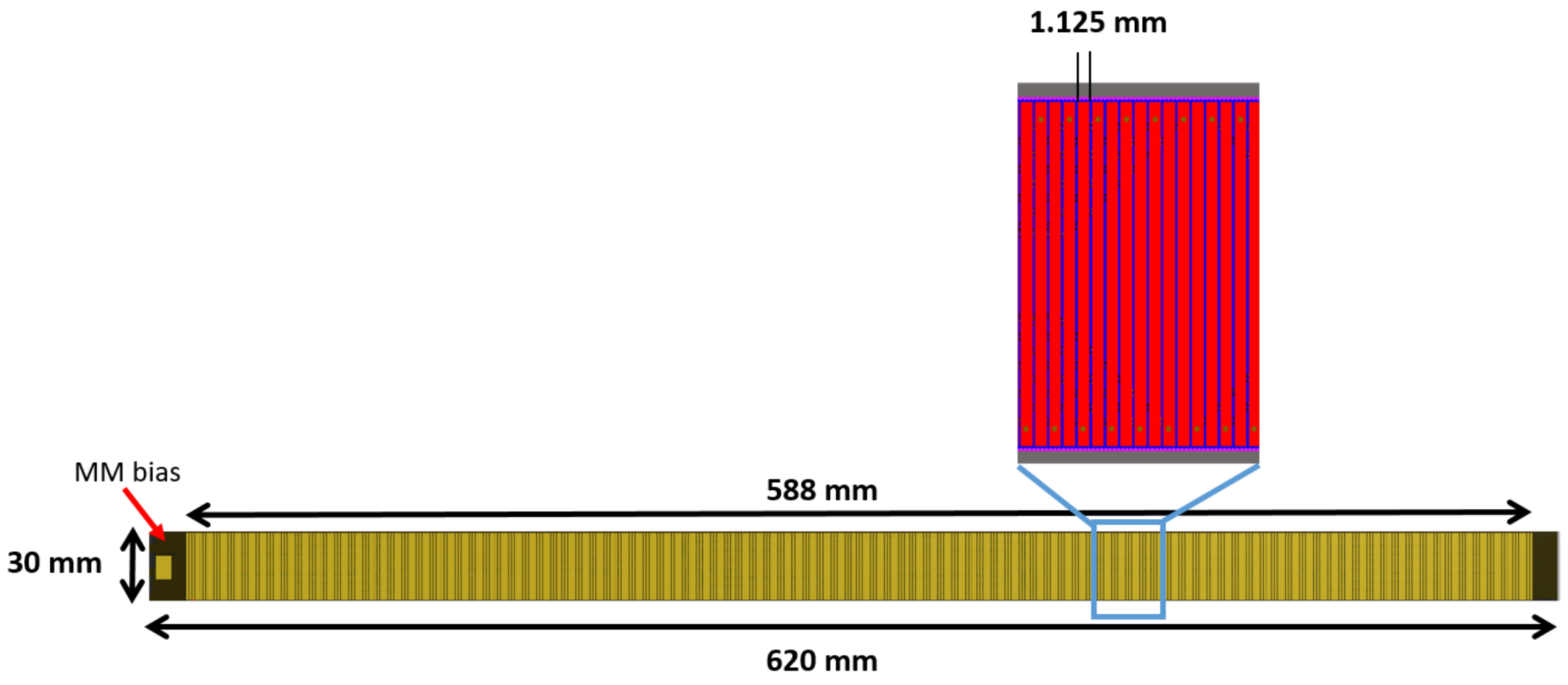

**Figure 3.** Schematic drawing of the position-sensitive Micromegas board.

A two-layer M-THGEM is used as a pre-amplification stage on top of the Micromegas. The gas gap between the Micromegas and the M-THGEM is of 3 mm. A constant electric field of 300 V/cm is kept between M-THGEM bottom electrode and the Micromegas mesh.

The M-THGEM is a novel hole-type avalanche structure consisting of a densely perforated assembly of multiple insulators (FR-4) substrate sheets, sandwiched between thin metallic (Cu/Au) electrode layers. The M-THGEM is produced by multi-layer printed-circuit-board (PCB) technique based on mechanical drilling the alternate layers of copper and core material laminated together. A small rim (typically 100 μm) is etched around the holes to prevent discharges due to mechanical defects. Each insulating layer has a thickness of 0.4 mm while the holes pitch and diameter are 1 mm and 0.5 mm, respectively. The effective area of the M-THGEM matches the effective area of the Micromegas boards. A suitable PCB via provides access for the voltage bias of the inner electrode. The M-THGEM can also be used as an active gate to reduce ion backflow effects in the drift region, as well as to suppress uncorrelated signals. This is achieved by selectively pulsing the bias of the M-THGEM top surface.

The configuration of the high-voltage bias of the various multiplication stages (M-THGEM and Micromegas) defines the gas gain, and may be adjusted depending on the charge and energy of the measured nuclei.

Two thin (~0.1 mg/cm^2) polypropylene foils with 5 mm thin conducting strips (made of 150 nm thick evaporated gold) seals the detector volume. The strips are biased through a resistive chain that creates a constant drop of voltage across the strips, so that a uniform electric field is kept across the drift region – typically around 18-20 Volt/cm.

In the present work, the detector volume is filled with a gas mixture consisting of 80% $CF_4$ and 20% of $iC_4H_{10}$ at a pressure of 40 torr. A typical gas flow rate of 5-10 sccm is constantly maintained and controlled by a mass flow controller. These were the same operational conditions that are used for the previous CRDC detector. In particular, the $CF_4/iC_4H_{10}$ gas mixture was chosen for the CRDC operation because of the high-drift velocity, high primary ionization, low electron diffusion and resistance to aging [7].

### 2.1 The front-end electronics and integration to the NSCL data acquisition system

The readout of the charge collected on the pads of the detector is performed by the front-end electronics of the GET system [8], developed for applications where a large number of channels are needed. The GET electronics are based on analog memories (Switch Capacitor Array) with a depth of 512 cells that store the waveforms in a circular buffer until a trigger is emitted to initiate the readout sequence. Several parameters, including preamplifier gain, shaping time, sampling frequency and depth of circular buffer can be configured via slow control. This offers great flexibility and easy adaptation to various detector configurations. Because the GET system was designed for TPC detectors, the range of sampling frequency between 3 and 100 MHz is well suited to the drift chamber readout. In addition, the back-end modules that receives the serialized data from the front-end has the capability to filter out unwanted waveforms and/or baseline data in order to reduce the readout dead time, as well as the overall amount of data transmitted to the network.

The 480 pads of the new tracking detector were connected to two AsAD (Asic and Adc) front-end electronic boards through ZAP intermediate boards (one per AsAD). Each AsAD board hosts four AGET (ASIC for GET) chips, along with a 4-channel 12-bit ADC (one channel per AGET) [9]. For each AGET board, 60 of the 64 available channels were used to process pad signals, thus covering the total 480 pads (60 x 4=240 pads per AsAD). The digital output from the two AsAD cards was sent to one CoBo (Concentration Board), mounted on a micro TCA crate. Among other things, the CoBo was responsible for timestamping data sent by each AsAd card and serving as a communication intermediary with the external network.

Figure 4 shows a sample of various waveforms recorded from a single event in the drift chamber. Compared to the old digital electronics that was based on the STAR TPC front-end boards, the GET electronics offers several advantages that boosts the performance of the drift chamber. Because it is based on modern ADC and FPGA technologies, the readout dead time of the GET system is smaller by more than an order of magnitude and can be easily optimized to match the dead time of the detector. In addition, at high beam intensities, piled-up signals on the same pad can be disentangled by analyzing the multiple peaks that appear in the recorded waveforms, resulting in a significant increase of the overall rate of the detector. This also opens the possibility to conceive experiments in which two or more particles transmitted through the spectrometer can be detected and tracked simultaneously by the drift chambers and associated detectors.

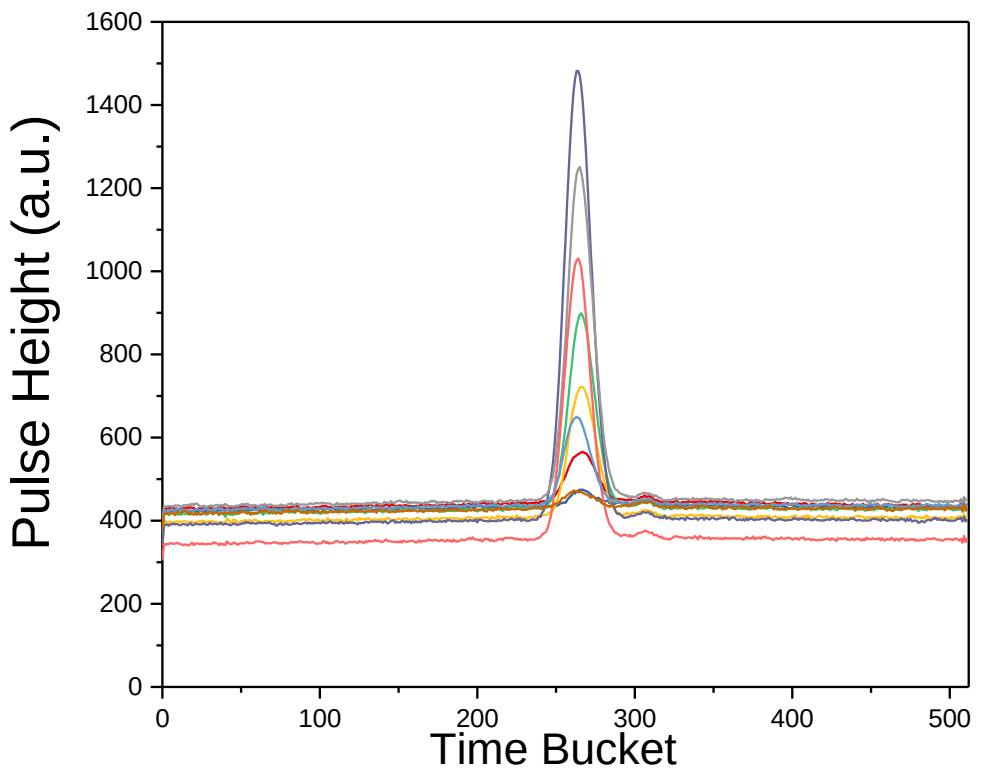

**Figure 4.** Signals recorded in various pads during an event.

In order to combine the information from the new drift chamber into the data stream from the rest of the S800 detectors, the GET electronics was integrated into the NSCL data acquisition system (NSCLDAQ). The NSCLDAQ is a modular toolkit for distributed data storing and data processing suitable for experimental nuclear physics. In this framework, an event is made of a set of time-stamped data sent from a so-called data source. Downstream from the data source, the data pipeline can include a series of "filters" to perform integrity checks and even pre-processing of data. This feature is critical, for instance, to reduce the data volume sent through the data pipeline, as discussed later. In addition, data events from different data sources can be combined, based on their timestamps, into a "merged" event using an Event Builder (EVB). The NSCLDAQ flexibility allows one to store and analyze data at any stage of the data pipeline.

In the NSCLDAQ framework, the GET electronics consisted of two data sources (one per AsAD). Data events from these two sources went through a filter (HitMaker) before being merged into a single GET event (more details in next section). The combination of the GET electronics into the S800 NSCLDAQ required an upgrade of the CoBo firmware to accept external clock and time-reset signals; this upgrade was crucial to enable the merging of both GET and S800 DAQ systems. Thus, common 10-MHz clock and reset signals were sent to the S800 and GET data sources to timestamp their corresponding data. A "master" EVB, at the end of the data pipeline, merged S800 and GET data events according to their timestamps. The trigger of the entire system was provided by a large plastic scintillator at the S800 focal plane. These signals went to an S800 trigger box which accepted the busy signal from the GET electronics and delivered an external "live" trigger to the CoBo.

### 2.2 Signal processing and position determination

The capability of the GET electronics to digitize the waveform of each pad signal provides a powerful tool to diagnose the response of the detector to different types of interacting particles. On the other hand, the analysis of such amount of data is prohibitively expensive, in terms of bandwidth (through the data pipeline), for the trigger rates of a regular experiment (few KHz) and the high granularity of the new detector. Therefore, the digitized waveforms from the GET electronics were pre-processed by the NSCLDAQ filter HitMaker. The purpose of this filter was to reduce the data bandwidth by extracting the "useful" parameters characterizing the waveform, before sending them down through the data pipeline instead of the entire digitized waveform.

Specifically, the signal amplitude (pulse height), charge (pulse integral), and time (pulse centroid) were extracted using a Lagrange interpolation (with up to 4 points interpolation), after subtracting the baseline. This condensed data was then sent down to the data pipeline, merged with the rest of the S800 data stream, and recorded for offline analysis. Note that the flexibility of NSCLDAQ enables analyzing (on-line) and recording of data from any stage through the data pipeline. As an example, the waveforms of Fig. 4 were recorded before going through the HitMaker filter. Two of the three waveform parameters (pulse height and centroid) were used to extract the x andy positions of the impinging particles. The x position (along the dispersive coordinate) is computed as the “center of gravity” (c.o.g.) of the pulse-height distribution projected over the anode pads. The position along the non-dispersive coordinate (y-coordinate) was derived from the pulse centroid, namely from the arrival time of the primary electrons created in the drift space with respect to an external trigger. Alternatively, the y-coordinate can be inferred by measuring the time properties of the signal sensed by the Micromegas mesh using a fast current preamplifier with respect to the external event trigger, as presently done with the CRDC detector.

## 3. Results

### 3.1 Performance evaluation with low-energy alpha-particles

A first performance evaluation was carried out by irradiating the detector with a low-energy alpha-particle source in a stand-alone configuration. Figure 5 depicts an energy spectrum (up to 8 MeV) of a 228-Th source recorded from the Micromegas mesh electrode using a charge sensitive preamplifier (Ortec Model 142 PC). The output signals from the preamplifier were processed by a linear amplifier (Ortec Model 572A) and a multichannel analyzer (Amptek Model Easy-MCA).

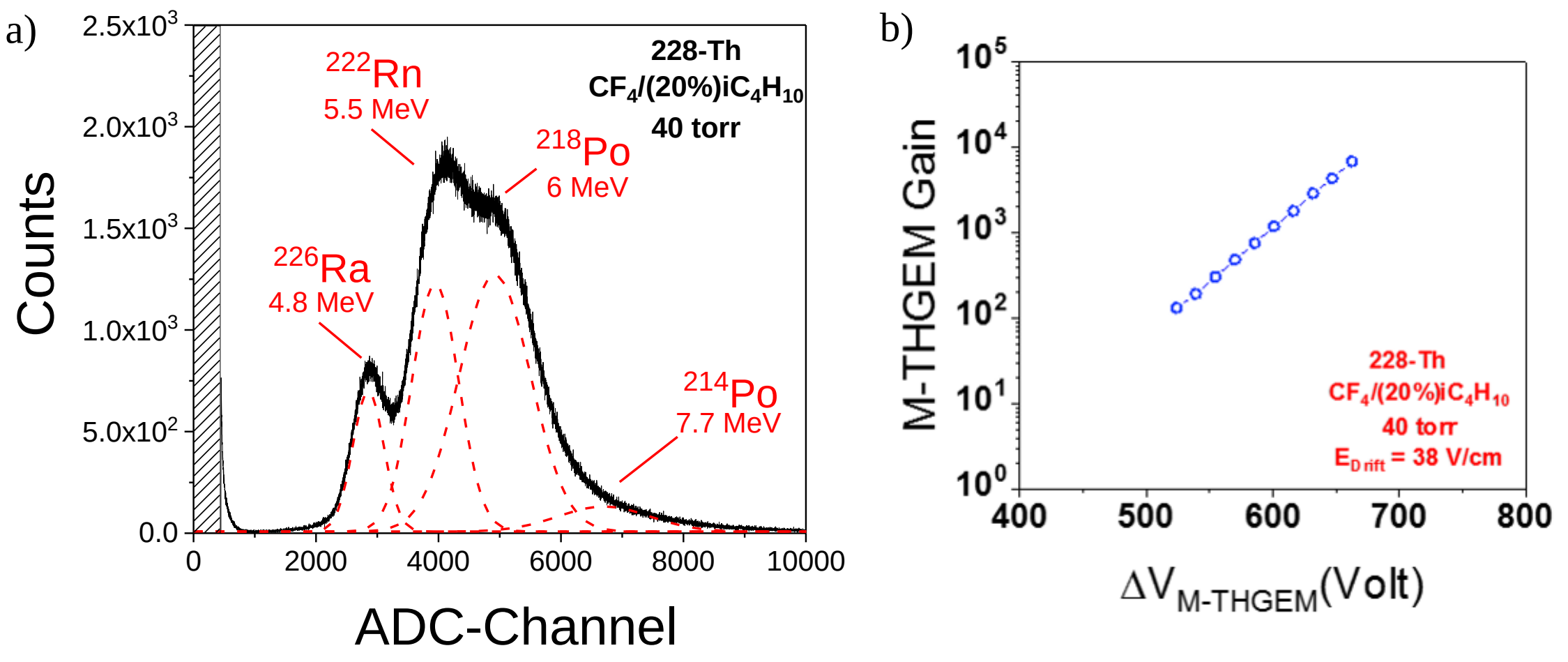

**Figure 5.** Energy spectrum of 228-Th recorded by the Micromegas mesh (part a). Gas gain provided by the M-THGEM pre-amplification element as function of the voltage different applied across each multiplication stage (part b). In both case, the detector operated in $CH_4$/20%i$C_4H_{10}$ at 40 torr (part a). Gas gain of the

The 228-Th source was enclosed in a holder equipped with a large aperture and placed at a distance of around 10 cm from the detector. In this configuration, the alpha particles impinge on the detector entrance window within a broad range of incident angles, corresponding to very different path-lengths across the detector effective volume. As a result, the recorded 5.5 MeV alpha-particle peak recorded in the energy spectrum has large fluctuations caused by the wide range of energy released by the alpha particles in the detector gas. This effect is clearly visible in the pulse-height distribution as a function of the position along with the Micromegas readout, as a sizable increase of signal amplitudes for those events corresponding to large incident angles (shown in Figure 6a).

A significant reduction of the angular straggling in the energy spectrum was accomplished by placing a (2.5 cm diameter) Silicon surface barrier detector (Si-b) downstream the DC, at a distance of 20 cm from the source, and selecting only those events that are simultaneously recorded by both the DC and the Si-b detector. Accordingly, only alpha-particles impinging on the DC surface within a small entrance angle (< 7 msr) are recorded. Under this irradiation condition, more similar to the one that characterizes the S800 focal plane detector system with an acceptance of 20 msr, the intrinsic energy resolution of the detector is estimated to be of the order of 20% (FWHM) for 4.8 MeV. At such a resolution, the peak of the high-energy alpha-particle component (8.7 MeV) and the low-energy ones (below 6 MeV) emitted by the 228-Th source can be clearly resolved – see the pulse-height mapping recording by the Micromegas readout in Figure 6b.

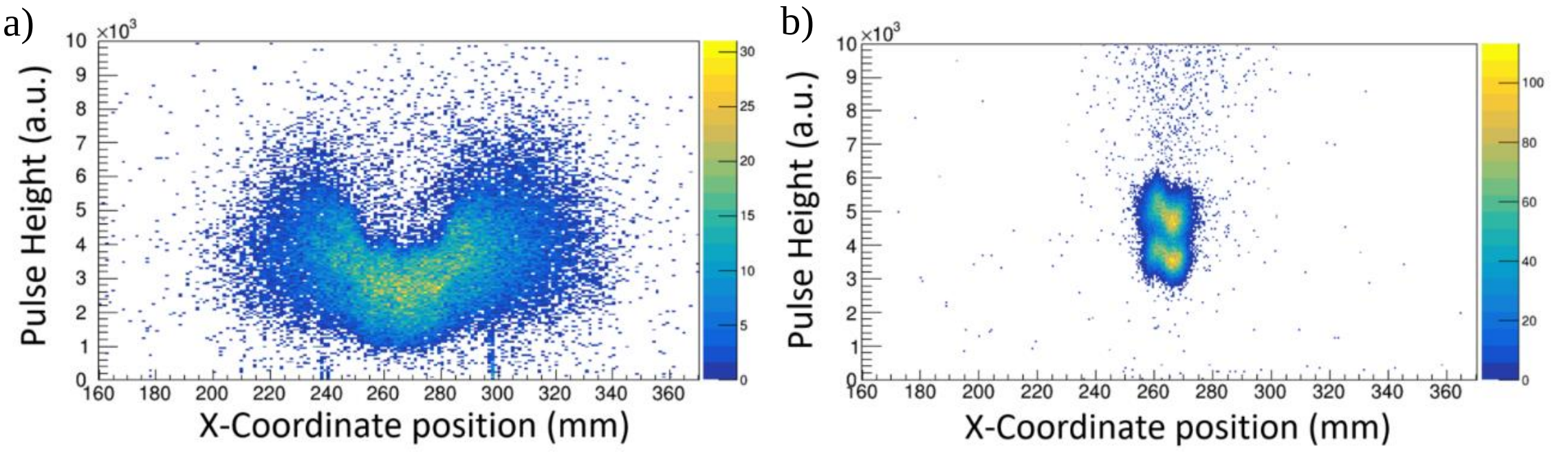

**Figure 6.** Mapping of the pulse height distribution as a function of the particle position along the Micromegas readout for a non-collimated source (part a) and for particles impinging on the narrow selected detector area by using a small Si detector as trigger (part b).

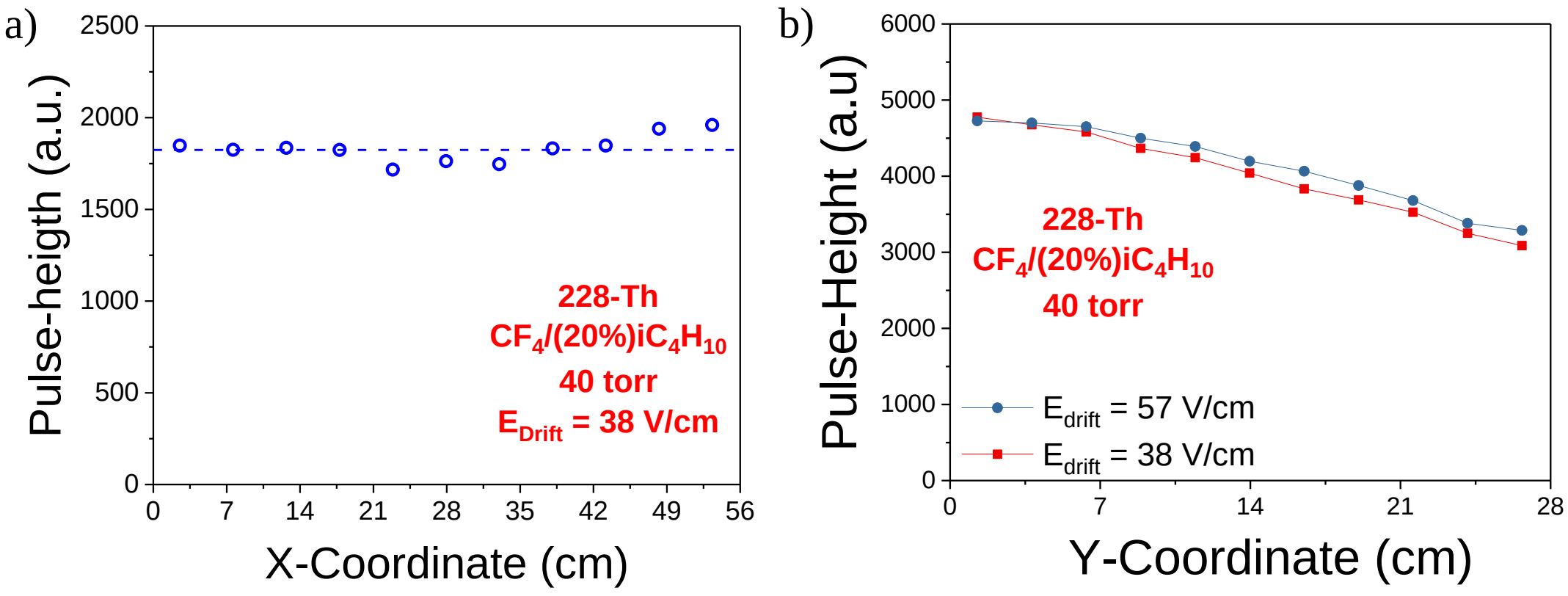

**Figure 7.** Pulse height variation across the gas amplifier structure boards (x-coordinate) and across the drift field region (y-coordinate).

One of the major sources of loss of energy resolution at low energies (below 10 MeV) is the large straggling caused by the thick entrance window foils. The DC is expected to provide a much better resolution at the S800 spectrometer focal plane, where particles at much higher energy (above hundred MeV) are generally detected.

Other contributions to the degradation of energy resolution include electronic noise, gain inhomogeneity across the readout boards (estimated to be of the order of ± 4%, see Figure 7a), and loss of primary electrons in the drift region. This includes electrons attachment as well as the electron loss on the field-cage foils due to the transversal diffusion on a narrow drift region (3 cm). As shown in Figure 7b, the average pulse heights measured when a strongly collimated source is irradiating at the top of the drift region is 30% lower compared to the irradiation at the bottom of the drift region. Variation of the pulse-height and energy resolution also affects the response of the detector in terms of position resolution and its uniformity.

### 3.2 Performance evaluation with high-energy heavy ion beams

A test of the tracking capabilities of the new detector readout has been performed at the S800 spectrometer focal plane using a 150MeV/u $^{78}Kr^{36+}$ beam and a cocktail beam of rare isotopes. The latter comprises isotopes with an atomic number Z from 4 to 36, produced as a secondary beam from the $^{78}Kr$ isotope impinging on a 2.7 mm thick Beryllium target. The DC was installed at the S800 focal plane in place of the downstream CRDC, just in front of the ionization chamber (Figure 8).

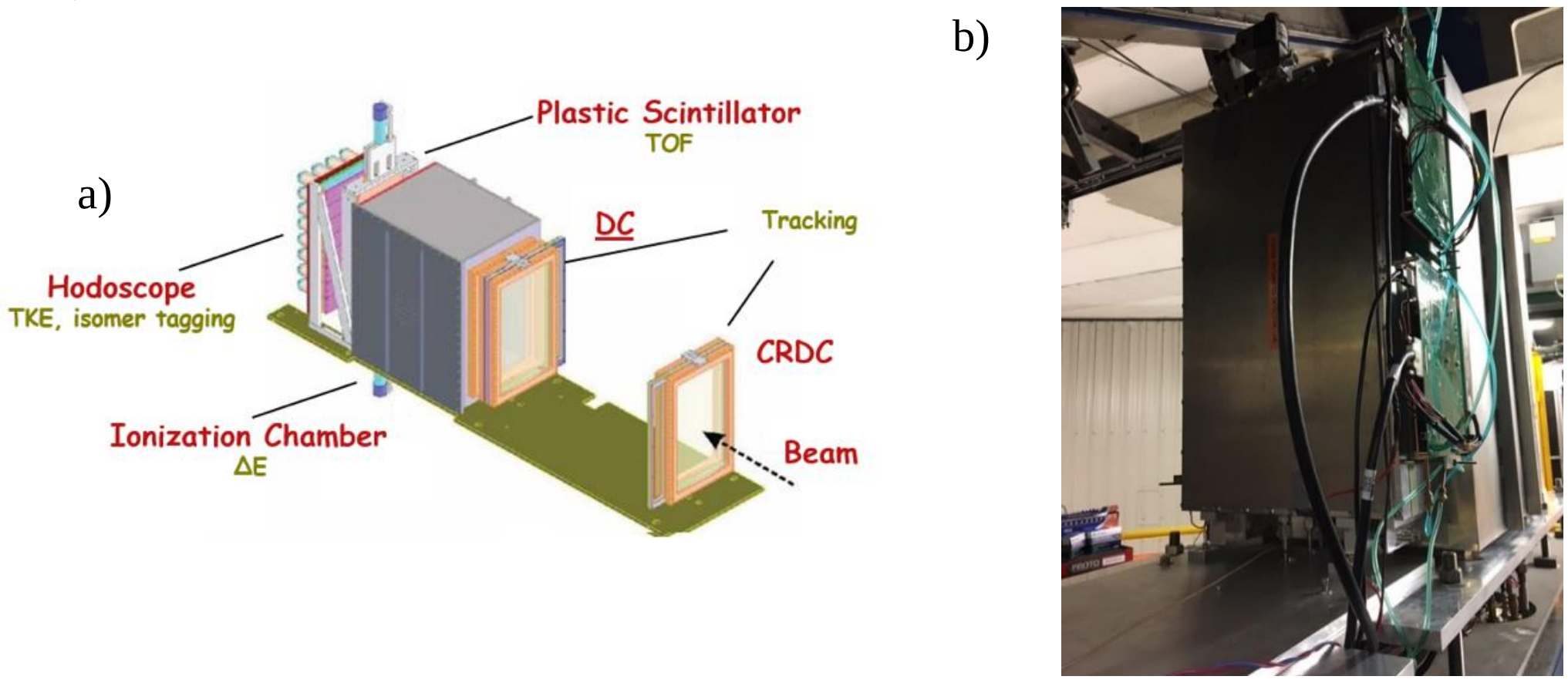

**Figure 8.** Schematic drawing of the detector configuration of the S800 focal plane used for tracking and particle identification (part a) and photograph of the new DC mounted in the s800 focal plan in front of the ionization chamber.

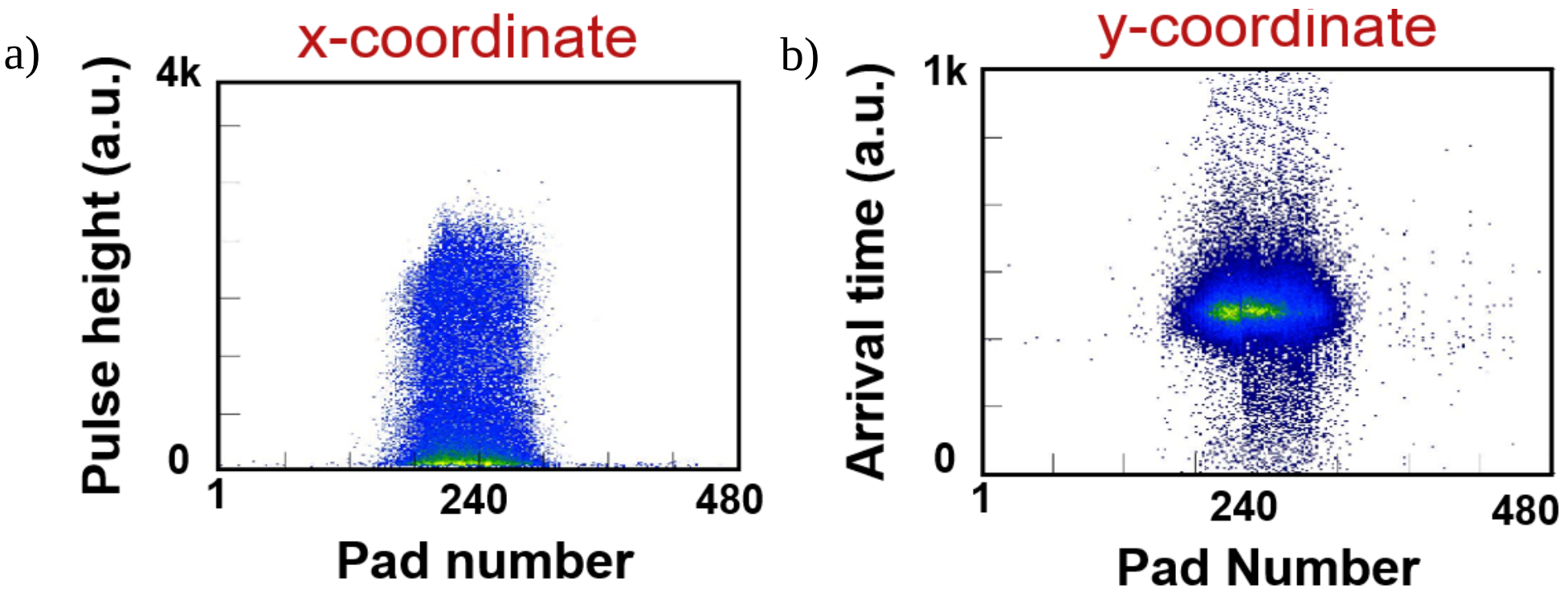

**Figure 9.** Cumulative mapping of the pulse-heights (x-coordinate, part a) and of the signal arrival time (y-coordinate, part b) measured along the Micromegas readout board (pads number) with the 150 MeV/u $^{78}Kr$ beam.

Figure 9 illustrate the cumulative mapping of the pulse-heights (part a) and of the drift time (part b), both inferred from the analysis of the pad waveforms of around 10k events, each of them corresponding to a $^{78}$Kr beam particle hitting the center region of the DC effective area.

As mentioned earlier, the processing of the distribution of pulse heights (Figure 9a) and of the drift times (Figure 9b), measured on an event by event basis, is used to compute the position along the x- and y-cordinate, respectivelly. An example of an image of a calibration mask placed upstream of the DC, recorded by irradiating a large area of the detector with the $^{78}$Kr beam, is shown in figure 10a. The holes in the mask have a diameter of 2.5 mm. The detector shows a good uniformity and a linear response across the whole effective area. Figure 10b shows the position spectrum obtained by analyzing the image of a hole from the calibration mask data of figure 10a. A position resolution of the order of 0.25 mm (sigma) was demonstrated along the dispersive coordinate (x-coordinate). A similar position resolution (below 0.5 mm - sigma) was obtained along the non-dispersive, y-coordinate.

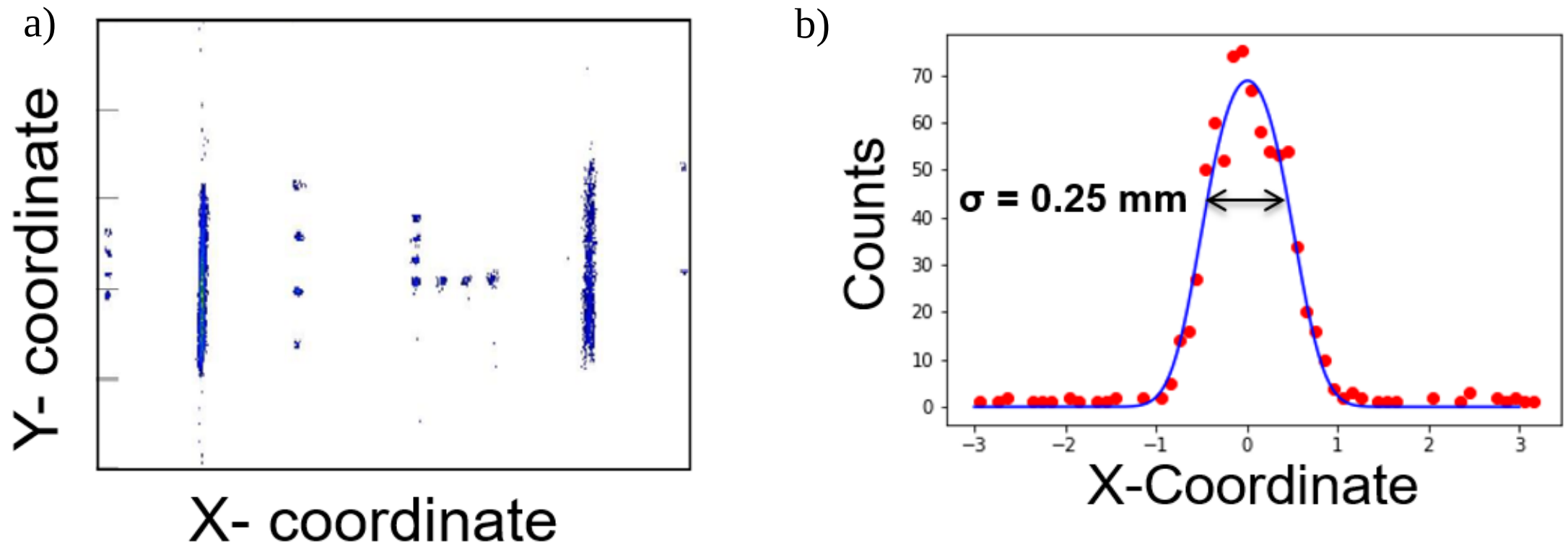

**Figure 10.** Part a: image of the calibration mask obtained by irradiating a large portion of the DC detector area with 150 MeV $^{78}$Kr. Part b: position resolution (0.25 mm sigma) estimated from the processing of a hole in the calibration mask image of part a).

The tracking capabilities of the drift chamber were also evaluated using low- to intermediate-heavy ion reaction fragments produced by the $^{78}$Kr beam impinging on a 2.7 mm thick beryllium target. The fragments transmitted to the S800 focal plane were uniquely identified on an event-by-event basis by combining the energy-loss, measured by the focal plane ionization chamber, and the time-of-flight (ToF) measurement performed using a beam-line timing detector upstream the target and the focal plane scintillator. Example of the PID for the $^{78}$Kr/Be fragment cocktail is shown in Figure 12.

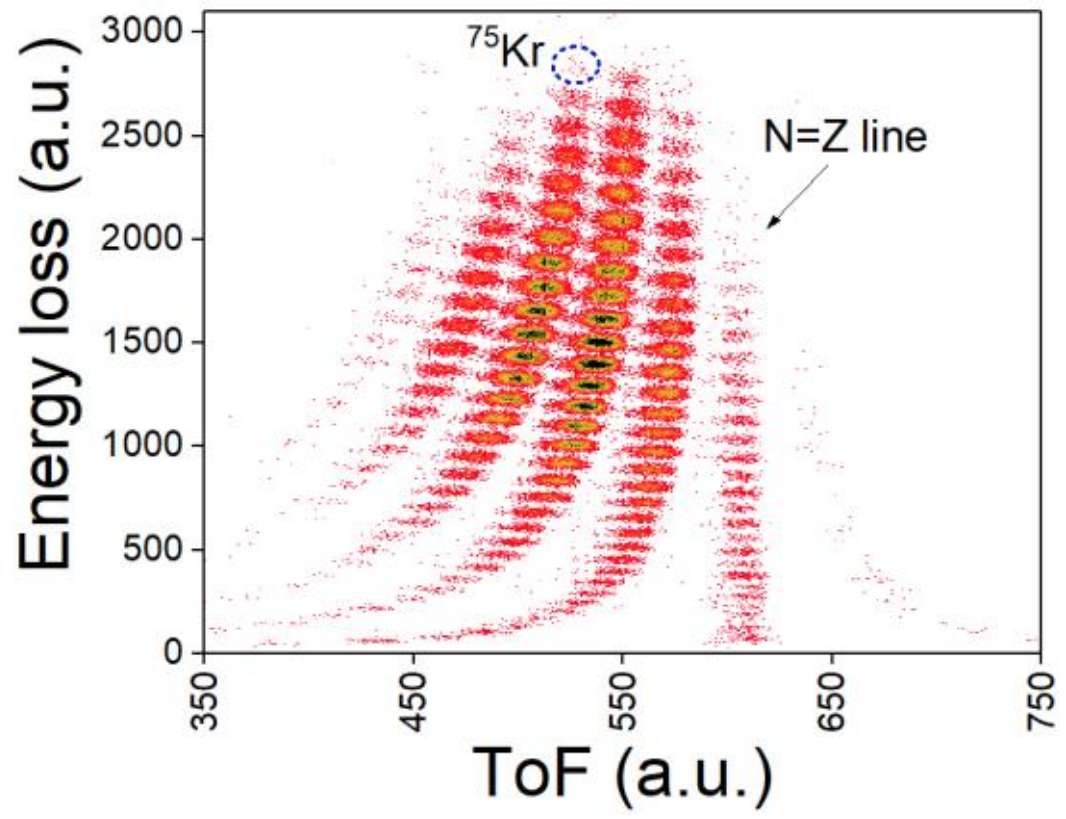

**Figure 11.** Energy loss (dE/E) as a function of the ToF for the $^{78}$Kr fragment cocktail transmitted to the S800 focal plane for the drift chamber detection efficiency test.

Besides the momentum vector calculation, a precise tracking capability over a wide range of transmitted nucleus species is crucial for improving the ToF resolution. This can be achieved by correcting the ToF dependence on the magnetic rigidity by using the transversal position in the dispersive image plane. One important feature of the new MPGD-based readout is the possibility to vary the gas gain by adjusting the voltage bias of the M-THGEM pre-amplification stage, enabling a full detection efficiency over a wide range in energy losses and a large dynamic range for the correction of the ToF measurements.

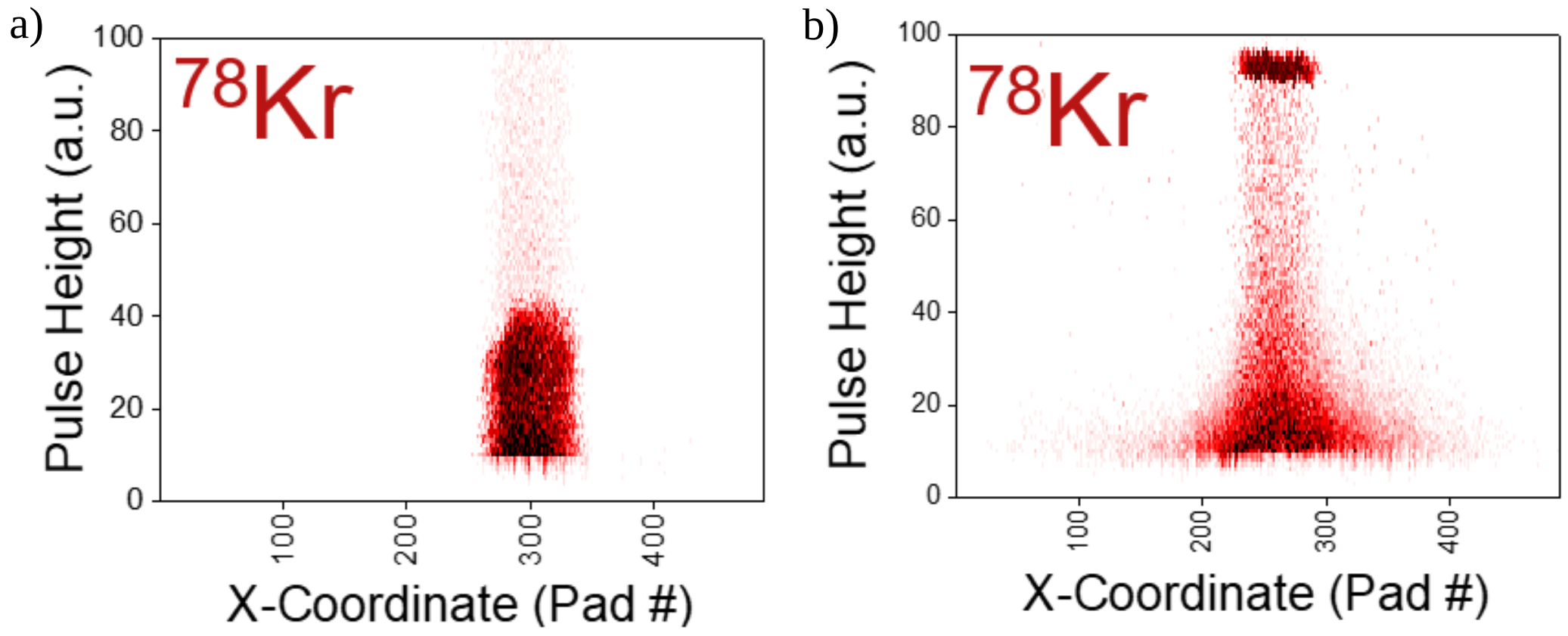

**Figure 12.** Cumulative mappings of the pulse heights for the 78Kr ion measured at low gas gain (part a) and at high gas gain (part b).

Figure 12 illustrates examples of pulse height mapping recorded for a set of selected fragments (including Beryllium, Oxygen, Potassium, Zinc, Chromium, and Krypton), obtained by operating the detector at a low gas gain (Figure 12a) and at a high gas gain (Figure 12b), respectively. In the former case, full detection efficiency and good tracking capabilities are obtained for elements heavier than Neon (see Figure 13). In contrast, lighter elements deposit less energy in the detector volume, so that most of the corresponding signals are below the detection threshold. This leads to a significant loss of efficiency. However, full detection efficiency for the lighter elements can be achieved at a higher gas gain, at the expense of saturating most of the signals corresponding to heavier elements. The localization and the track information with saturated signals along the readout plane can be restored by implementing a more elaborated data processing algorithm, which involves fitting the non-saturated tails of the pulse-height distribution. Various dedicated algorithms for GET-based DAQ have been already demonstrated (see for instance [9]), though this implies an acceptable loss of localization accuracy (estimated to be up to 0.5 mm sigma).

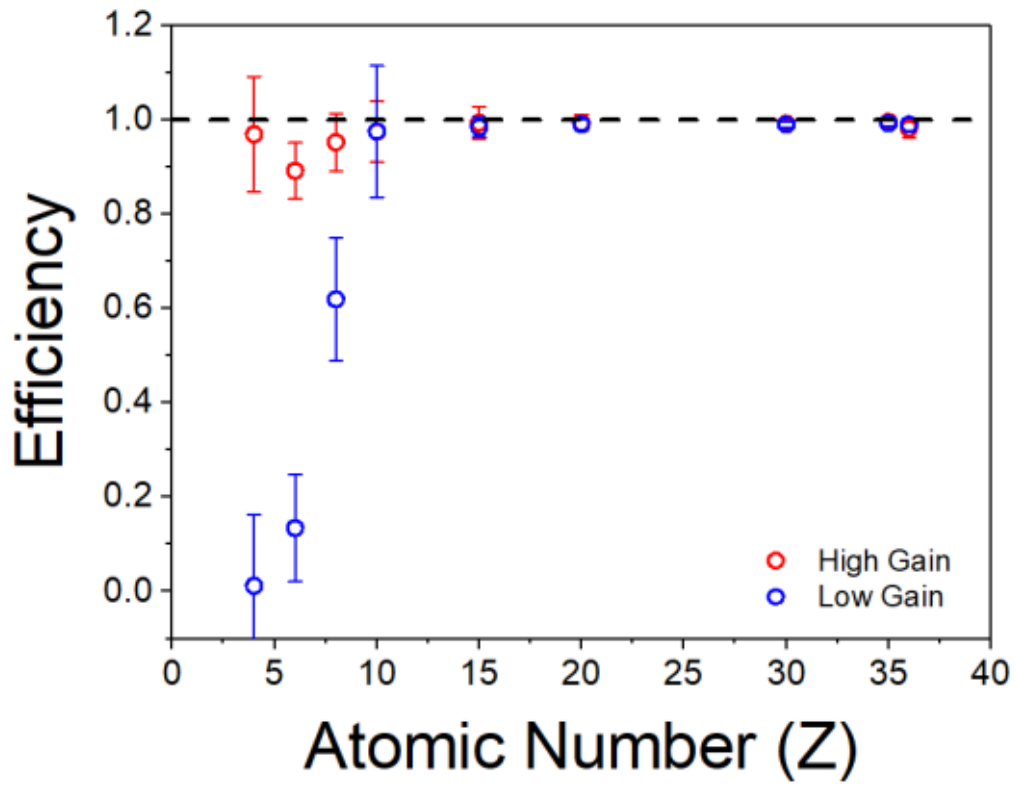

**Figure 13.** Detection efficiency of the drift chamber as a function of the Z-number for low (red graph) and high (blue graph) gas gain operation.

## 4. Summary and discussion

We report on the development of a new drift chamber readout concept planned for the upgrade of the S800 spectrometer focal plane detector system. The new detector design uses a new simple and robust gas avalanche readout scheme based on micro-pattern gaseous detector technology combined with a new DAQ hardware based on the GET electronics.

The hybrid combination of position-sensitive Micromegas board and the multi-layer M-THGEM, used as a pre-amplification stage, provides stable high-gain operation at low pressure (below 50 torr), suitable for heavy-ion tracking. This configuration has the unique feature of sharing the total gas gain between a cascade of elements, each one operated below the critical voltage for discharges. In addition, it also provides a better ion backflow reduction which leads to a large improvement in long-term free-spark stability. This hybrid MPGDs configuration, characterized by a gas amplification region (M-THGEM) separated from the readout stage (Micromegas) [7], shows a remarkably stable high-gain operation, as well as negligible aging effects. The high radiation hardness of the MPGDs allows the new readout scheme to be operated in faster gas mixtures (i.e. using argon based mixtures), instead of the conventional $CF_4$/isobutane mixture used for the CRDC. Furthermore, the new digital DAQ based on GET electronics provides the possibility to process multi-hit events. Fast filling gas and multi-hit capability allow for an overall increase of the counting rate capability, up to 20 kHz compared to the present limit of 5 kHz.

The higher granularity of the readout board and the fast response of the MPGD architecture provide a spatial resolution (sigma) of 0.25 mm and below 0.5 mm in the dispersive and non-dispersive coordinate, respectively.

Compared to the old digital electronics that was based on the STAR TPC front-end boards, used for the CRDC readout, the GET system offers several advantages that boost the performance of the drift chamber. Because it is based on modern ADC and FPGA technologies, the readout dead time of the GET system is smaller by more than one order of magnitude and can be easily optimized to match the dead time of the detector. In addition, the possibility to record several pulses on each channel means that pile-up effects can be corrected for and the overall rate on the detector can be increased significantly. These features also open the possibility to conceive experiments in which two or more particles transmitted through the spectrometer can be detected and tracked simultaneously by the drift chambers and associated detectors.

The successful development of the new tracking system for the S800 spectrometer will allow to carry out the rich program of nuclear physics with exotic beams at the present NSCL and at the future FRIB, and would also serve as a prototype for a similar, bigger tacking system needed for FRIB's High Rigidity Spectrometer, currently in construction.

## Acknowledgment

This material is based upon work supported by the National Science Foundation under Grant No. PHY-1565546. The authors are particularly grateful to Andrea Stolz and John Yurkon for their valuable and constructive suggestions during the planning and development of this research work. Assistance given by the NSCL A1900 fragment separator group during the beam test has been a great help and much appreciated.